\documentclass[pdftex]{aa}
\usepackage[varg]{txfonts}
\usepackage{natbib}
\usepackage{graphicx}
\usepackage{float}
\usepackage{lscape}
\usepackage{caption}
\usepackage{subcaption}
\usepackage{sidecap}
\usepackage{xcolor}
\usepackage{comment}
\titlerunning{}

\begin{document} 

   \title{Shape of the outer stellar warp in the Large Magellanic Cloud disk}
   \author{Saroon S
          \inst{1,2}\fnmsep\thanks{saroonsasi19@gmail.com}
          \and
           Smitha Subramanian\inst{1}\fnmsep\thanks{smitha.subramanian@iiap.res.in}
          }
   \institute{$^{1}$ Indian Institute of Astrophysics, Koramangala II Block, Bangalore-560034, India \\ $^{2}$ Amrita Vishwa Vidhyapeetham, Amritapuri, Kollam , Kerala - 690525 , India\\
            } 
   \date{Received---- ; accepted----}
  \abstract
   {Warps are vertical distortions of the stellar or gaseous disks of galaxies. One of the proposed scenarios for the formation of warps involves tidal interactions among galaxies. A recent study identified a stellar warp in the outer regions of the south-western (SW) disk of the Large Magellanic Cloud (LMC) and suggested that it might have originated due to the tidal interaction between the LMC and the Small Magellanic Cloud (SMC). Due to the limited spatial coverage of the data, the authors could not investigate the counterpart of this warp in the north-eastern (NE) region, which is essential to understanding the global shape, nature, and origin of the outer LMC warp. In this work, we study the structure of the LMC disk using data on red clump stars from the \textit{Gaia} Early Data Release 3 (EDR3), which cover the entire Magellanic system. We detected a warp in the NE outer LMC disk which is deviated from the disk plane in the same direction as that of the SW outer warp, but with a lower amplitude. This suggests that the outer LMC disk has an asymmetric stellar warp, which is likely to be  a U-shaped warp. Our result provides an observational constraint to the theoretical models of the Magellanic system aimed at improving the understanding the LMC-SMC interaction history.}
\maketitle
\section{Introduction}
The outer parts of most of the disk galaxies ($\sim$ 60 $\%$, \citealt{2003sanchez}) are distorted with respect to the inner disk plane forming warps (vertical distortions), which are easily identifiable in an edge-on view. Although warps are mostly found in the atomic hydrogen gas disks (\citealp{1976A&A....53..159S}; \citealp{1978PhDT.......195B}B), stellar warps are also observed (\citealp{1981A&A....95..105V}).
 Different observed types of warps include: S-shaped  (integral-shaped) warps in which both sides are warped but in opposite directions; asymmetric warps with significantly different amplitudes on both sides; U-shaped warps in which both sides are warped in the same direction; and one-sided warps, also known as L-shaped warps  (\citealt{1990MNRAS.246..458S,1998A&A...337....9R,2003sanchez,2006NewA...11..293A}). The Milky Way, with its integral-shaped warp, was the first ever warped galaxy discovered (\citealp{1957AJ.....62...90B,1957AJ.....62...93K}), based on 21 cm hydrogen (HI)-line observations.  The subtle stellar warp of the Milky Way was also recently observed by various stellar surveys (\citealp{2018MNRAS.478.3809S,2019NatAs...3..320C,2019A&A...627A.150R,2019Sci...365..478S}). Stellar warps have smaller amplitudes  compared to HI-warps  (\citealp{1979A&AS...38...15V,1980ApJ...236L...1S,1987PASJ...39..849S}) and 70$\%$ of the stellar warps are (integral) S  shaped (\citealp{2006NewA...11..293A}). Many of the observed S-shaped warps are also found to be asymmetric (\citealt{garcia2002}). Even U-shaped and one-sided or L-shaped warps have also been observed (\citealp{2003sanchez}). \\
Theoretically, there are many mechanisms proposed for the formation and evolution of warps. Different scenarios include the misalignment of the galactic disk with the angular momentum of the halo  \citep{1999ApJ...513L.107D}, accretion of the intergalactic medium into the galactic disk (\citealp{2006MNRAS.370....2S}), and cosmic in-fall causing the outer parts of the halos to reorient  (\citealp{1999MNRAS.303L...7J,1989MNRAS.237..785O}).  Tidal interactions by nearby companions (\citealp{2001A&A...373..402S,1990A&A...233..333K}) and fly-by encounters (\citealp{2000ApJ...534..598V,Kim_2014}) have also been suggested to be responsible for the formation of warps. \citet{1998A&A...337....9R} showed that massive galaxies are less likely to warp and the most interacting galaxies show measurable warps, emphasising the role of gravitational interaction. \\
In this context, it is an interesting task to study the disk structure of one of the nearest interacting low-mass disk galaxies, namely, the Large Magellanic Cloud (LMC). The LMC is a disk galaxy with a planar geometry and is located at a distance of  $50\pm 2$ kpc (\citealp{de2014clustering}). It is interacting with its neighbour, the Small Magellanic Cloud (SMC). 
Previously, various authors (\citealp{van_der_Marel_2001, 2002AJ....124.2045O,2004ApJ...601..260N,2009AJ....138....1K,Subramanian2010,Subramanian2013,Inno2016} and references therein) had estimated the orientation parameters of the LMC disk using different tracers. Some studies (\citealt{2002AJ....124.2045O, Subramaniam2003, Subramanian2010,Subramanian2013}) have also identified warped and flared structures. As the LMC covers a large area in the sky, all of these studies were limited by the spatial coverage of the  data and hence restricted mainly to the inner LMC disk. \\
\begin{figure*}
\centering
     \begin{subfigure}[b]{0.34\textwidth}
         \includegraphics[width=\textwidth]{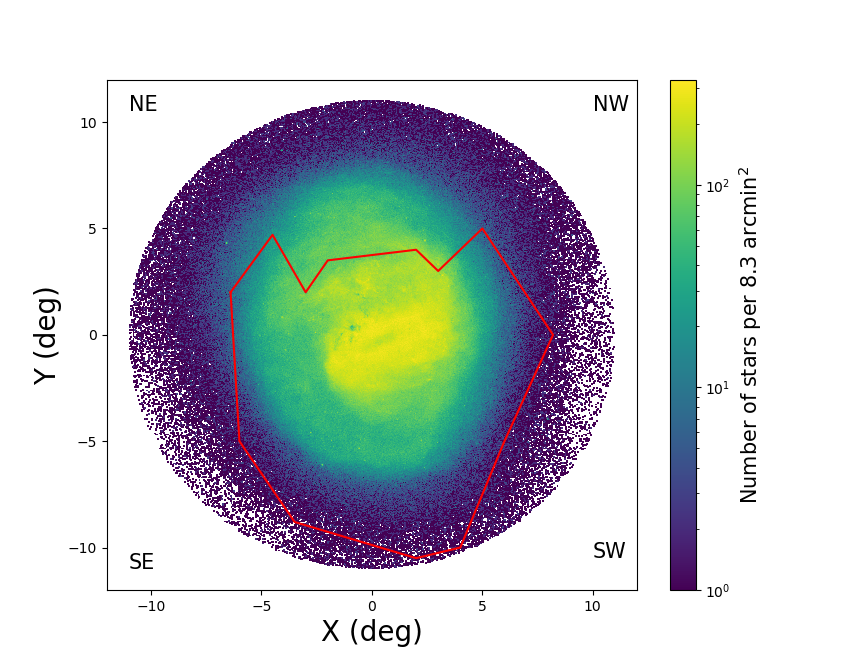}
     \end{subfigure}
      \begin{subfigure}[b]{0.30\textwidth}
         \includegraphics[width=\textwidth]{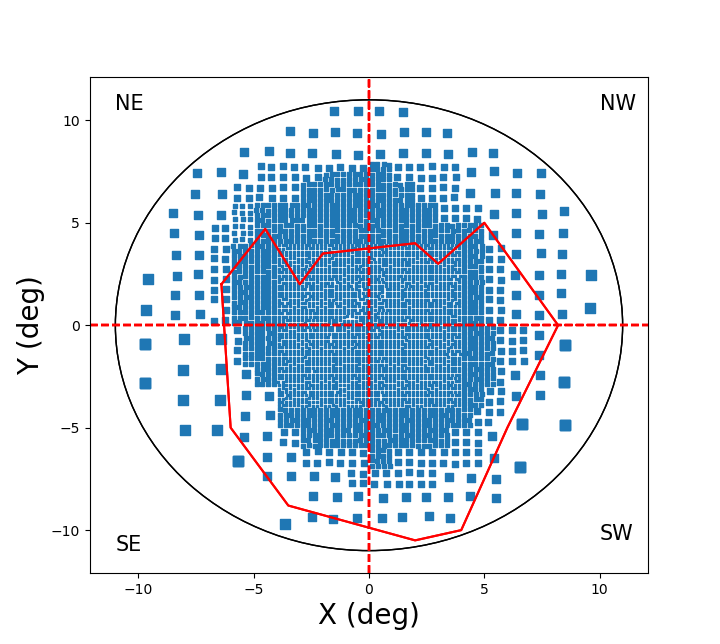}
     \end{subfigure}
     \begin{subfigure}[b]{0.35\textwidth}
         \includegraphics[width=\textwidth]{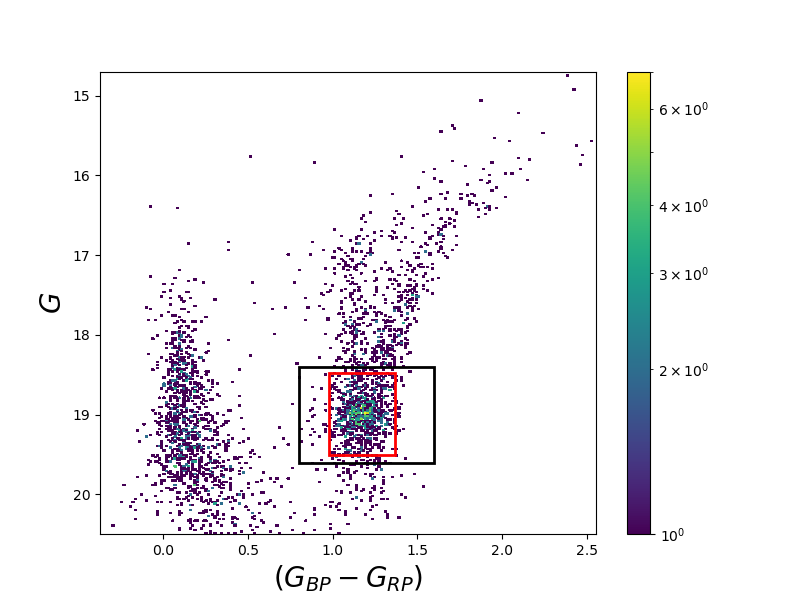}
     \end{subfigure}
        \caption{Spatial distribution of the LMC sources, the sub-regions analysed and the Hess diagram representing the stellar density in the observed CMD of an outer sub-region are shown. Left-hand panel shows the Cartesian plot with the sources from the \textit Early Data Release 3 (EDR3) for an area of $\sim$ 380 deg$^2$ of the LMC. The colour bar from blue to yellow represents the increase in stellar density (number of stars per bin of size 8.3 $arcmin^2$ , in units of $arcmin^{-2}$ )
        The middle panel illustrates the binning of the analysed LMC region into sub-regions based on the stellar number density. The X and Y are defined as in~\citealp{van_der_Marel_2001} with origin as LMC's photometric centre ($82^\circ.25,-69^\circ.45$) ;~\citealp{van_der_Marel_2014}. The east and north are towards left and up, respectively. The red dashed lines show the division of NE, NW, SE, and SW sub-regions. The blue squares show each sub-region represented by their central coordinates and the size of the square illustrates the increase in bin-size with the drop in stellar number density as we move towards the outskirts of the LMC disk. The outline of the region studied by \cite{choi2018smashing} is shown by the red polygon in the left and middle panels. Right-hand panel shows the Hess diagram representing the stellar density in the observed CMD of Gaia EDR3 sources enclosed in one of the outer sub-regions. The  black rectangular box shows the initial RC selection box and the red rectangular box is the redefined RC region.}
        \label{fig001}
\end{figure*}
A recent study by \citep{choi2018smashing} used the optical data of red clump (RC) stars from the Survey of the Magellanic Stellar History (SMASH,~\citealp{2017AJNidever}), which covers a larger area of the LMC than the previous studies, and found a prominent warp on the south-western (SW) region of the outer LMC disk. The study suggested that this feature might have formed due to the direct collision of the LMC and the SMC in the recent past. But due to the limited spatial coverage of the SMASH data, they could not study the entire LMC disk, especially the north-eastern (NE) part of the outer LMC disk to find the counter-part of the SW outer warp. They also suggested that the earlier identified extra-planar features are just part of the rippled inner disk and the derived global shape of the entire LMC disk is highly dependent on the spatial coverage of the analysed data. \\
In this study, we use the \textit{Gaia} Early Data Release 3 (EDR3, \citealp{2021A&AEDR3}) data, which covers the entire Magellanic System, to estimate the orientation parameters of the LMC disk and to find extra-planar features. The goal is mainly to look for the presence (absence) of a warp in the unexplored NE part of the outer LMC disk to get a deeper understanding of the global shape, nature, and origin of the outer warp of the LMC disk. For this study, we use the RC stars as tracers. The RC stars are more massive and metal-rich counterparts of the horizontal-branch stars, with an age and mass range of 2--9 Gyr and 1--2.2 M$_{\sun}$, respectively  (\citealt{2001MNRAS.323..109G,2016ARA&A..54...95G,Subramaian2017}). They start their core helium-burning phase at an almost-fixed core mass. Hence, they have fixed absolute magnitudes and serve as useful tracers to study the three dimensional (3D) structure of their host galaxies.
\section{Data}
We used photometric data from the \textit{Gaia} EDR3 \citep{2021A&AEDR3photo}  for our study. The left panel of Figure \ref{fig001}, centred on the photometric centre of the LMC, $82^\circ.25,-69^\circ.5$, (\citealp{van_der_Marel_2001} ;~\citealp{van_der_Marel_2014}) and having an area of $\sim$ 380 deg$^2$ of the LMC is taken for our analysis. This covers almost the entire LMC region with high stellar density, especially the unexplored NE part of the disk. The outline of the region studied by \cite{choi2018smashing} is shown by the red polygon on the left and middle panels of Figure \ref{fig001}.\\
A set of selection criteria are used to select the most probable LMC sources. 
To remove the foreground stars that lie below 5 kpc, a cut in parallax $(<= 0.2)$ is given. 
A 5-sigma deduction in flux over errors (G, $G_{BP}$ and $G_{RP}$) is applied to select only those sources with uncertainty $\le$ $20\%$ and applied a cut on colour excess factor as, $1.0+0.015(G_{BP} - G_{RP})^2 <$ phot\_bp\_rp\_excess\_factor $< 1.3+0.06(G_{BP} - G_{RP})^2$. In addition, we selected sources with astrometric excess noise $<=1.3$ mas and re-normalised unit weight error (RUWE) value of $<$ 1.2. 
Finally, we applied a cut to the proper motion values  based on the stellar density in the proper motion space (1 mas yr$^{-1}$ $\le$  $\mu_{\alpha}$  $\le$ 2.5 mas yr$^{-1}$ and $-$0.8 mas yr$^{-1}$ $\le$  $\mu_{\delta}$ $\le$ $+$1.5 mas yr$^{-1}$). 
This cut-off applied to the proper motion values further reduces the Milky Way contamination in the data.\\
The analysed region of the LMC covers the entire area where the RC feature can be easily detected in the colour magnitude diagram (CMD). The 380 deg$^2$ LMC region, consisting of about 6 million stars after the application of the selection criteria, is divided into 3626 sub-regions. The size of the sub-region is decided based on the visual identification of the RC region in the Hess diagram. We identified that a minimum of 150 stars in the RC selection box allows a clear identification of the RC region in the Hess diagram. Again, the RC region in the CMD has contamination from the Red Giant Branch (RGB) stars. Typical contamination from the RGB stars in the peak RC magnitude range is $\sim$ 20-30 percent. Thus a minimum number of 150 stars in the RC selection box suggests only $\sim$ 120-100 most probable RC stars, which sets a minimum of $\sim$ 10 $\sigma$ detection of most probable RC stars in each sub-region.
To obtain a minimum of 150 stars in the RC selection box,  we used a bin size of 10' $\times$ 10' $arcmin^2$ in the central parts of the LMC (upto 4$^\circ$) and in the outer regions, it varied from 20' $\times$ 20' to 120' $\times$ 120'$arcmin^2$ and is illustrated in the middle panel of Figure \ref{fig001}. We note that at the LMC distance, the largest sub-region has a size of $\sim$ 1.7 $\times$ 1.7 kpc$^2$. This is relatively large to identify stellar sub-structures that span a smaller area. But these larger sub-regions are mostly in the south-eastern part of the LMC where the stellar density is less. The NE part of the LMC, which is of main interest in this study, has higher stellar density and most of the sub-regions in the outer NE part have sizes of $\sim$ 0.28 $\times$ 0.28 kpc$^2$ (20 $\times$ 20 $arcmin^2$) and 0.43 $\times$ 0.43 kpc$^2$ (30 $\times$ 30 $arcmin^2$). This can be seen in the middle panel of Figure 1. However, the effect of the size of sub-regions on our final results are discussed in Section 5.
\section{Analysis}
\begin{figure*}
     \begin{subfigure}{0.5\textwidth}
         \includegraphics[width=\textwidth]{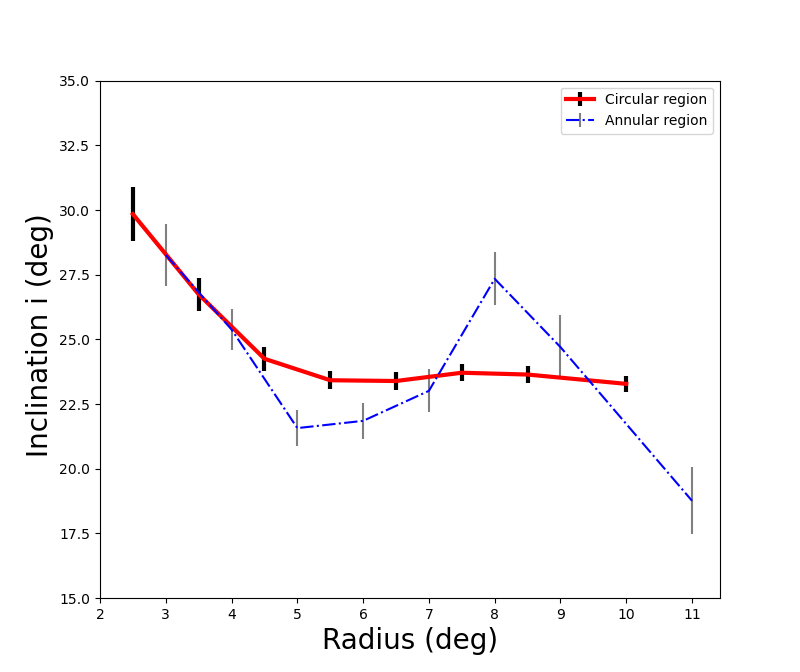}
     \end{subfigure}
     \begin{subfigure}{0.5\textwidth}
         \includegraphics[width=\textwidth]{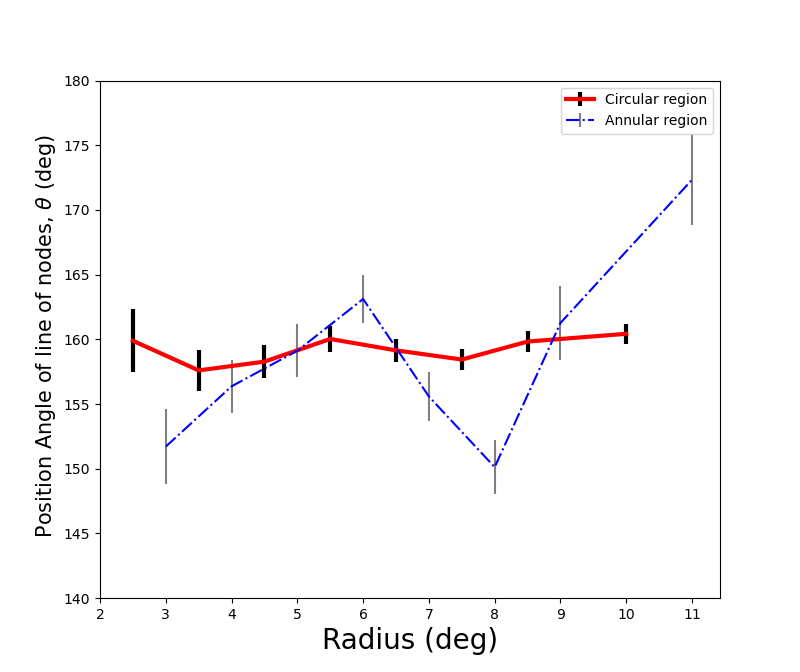}
     \end{subfigure}
        \caption{ Radial dependence of the measured inclination (left) and position angle of the line of nodes (right). The red solid line shows the variation of parameters with each circular radius and the blue dash-dot line corresponds to the variations with each annular radius. The black and grey vertical lines represent the error bars associated with each point.}
        \label{fig0033}
\end{figure*}
\subsection{Hess diagram and selection of RC stars}
We constructed Hess diagrams (stellar density plots) of the ${G}$ versus $(G_{BP} - G_{RP})$ CMD for all sub-regions, with bin sizes of $\sim$ 0.01 and 0.03 mag in $(G_{BP} - G_{RP})$ and in ${G,}$ respectively. 
 The right panel of Figure \ref{fig001} shows the Hess diagram of a sub-region in the  LMC disk. The RC stars in the sub-region are initially identified using the black rectangular box shown in the Hess diagram. The red rectangular box shows the re-defined or final RC selection box, as explained in Section 3.2. The colour bar from blue to yellow represents the increase in the number density of stars. Similar Hess diagrams for all the 3626 sub-regions are plotted and the RC stars in each sub-region are selected. The initial RC selection box is defined based on the visual identification of the RC region in the CMD of each sub-region. On average, in every sub-region, the RC stars lie within a range of $0.8 - 1.6$ mag in $(G_{BP}-G_{RP})$ and $18.2 - 19.8$ mag in $G$. 
\subsection{Median RC magnitudes} 
Initially, the median magnitude and colour, along with their standard deviation ($\sigma$) of RC stars (inside the initial RC selection box) in each sub-region were calculated. 
Based on these values, we redefined the RC selection box of all the sub-region. The size of this box is defined as median magnitude-colour $\pm$ 2$\sigma$ magnitude-colour and is indicated by the red box in the right panel of Figure \ref{fig001}. This helps to obtain a more accurate RC selection.
The median magnitude and the associated standard error of the RC stars inside the redefined selection box are obtained for all the 3626 sub-regions.
\begin{figure*}
     \begin{subfigure}{0.5\textwidth}
         \includegraphics[width=\textwidth]{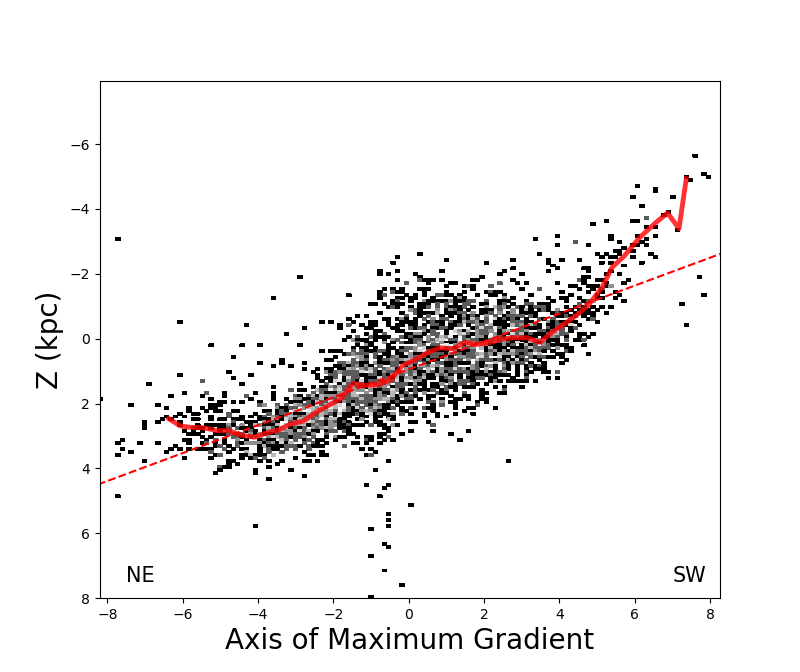}
     \end{subfigure}
     \begin{subfigure}{0.5\textwidth}
        \includegraphics[width=\textwidth]{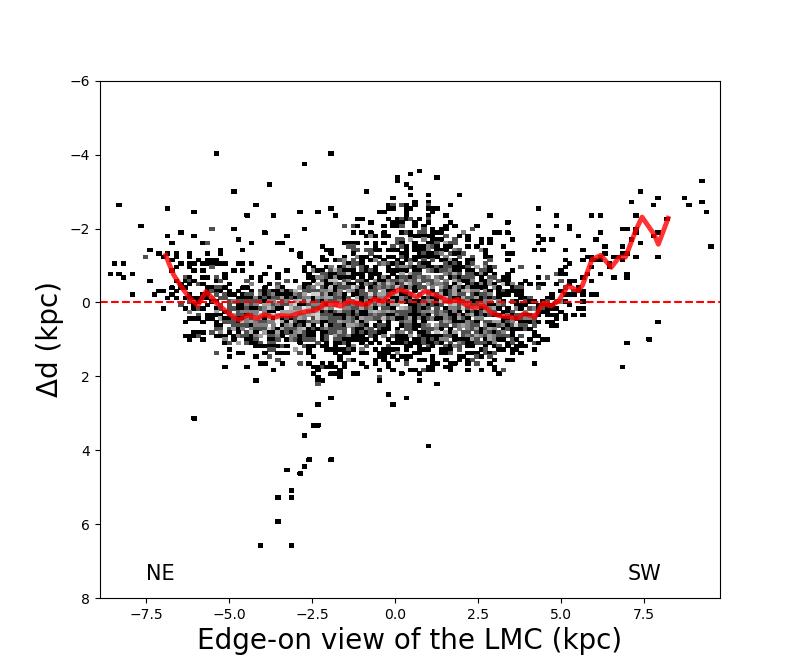}
     \end{subfigure}\\
     \begin{subfigure}{0.5\textwidth}
         \includegraphics[width=\textwidth]{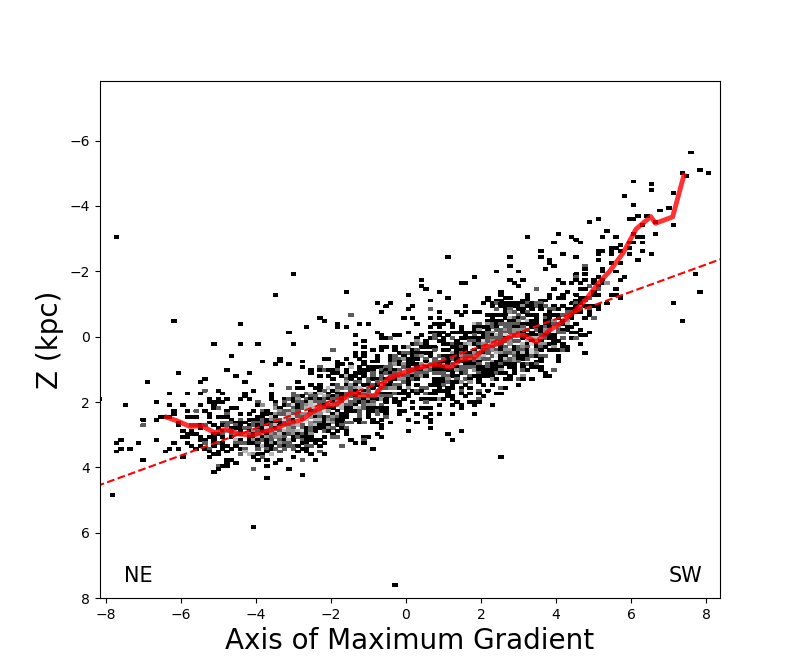}
     \end{subfigure}
     \begin{subfigure}{0.5\textwidth}
        \includegraphics[width=\textwidth]{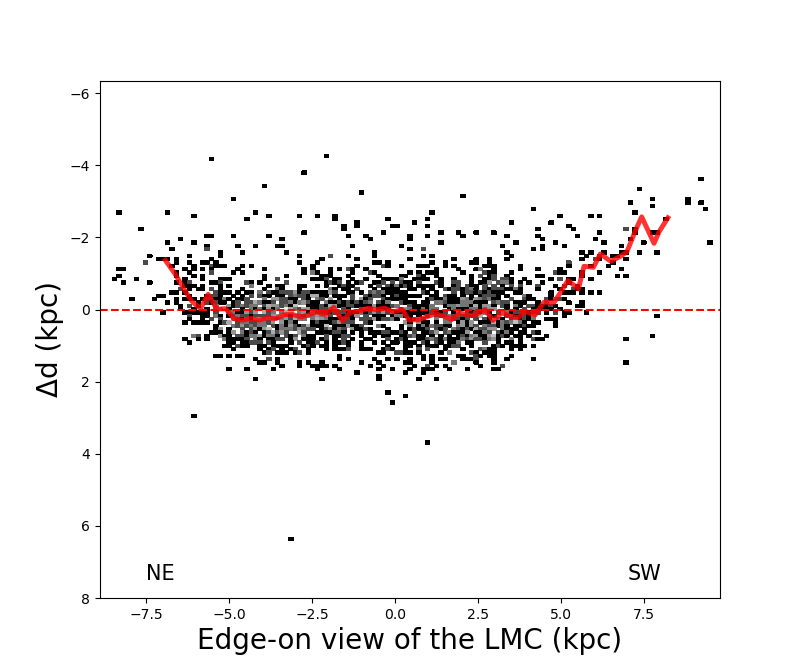}
     \end{subfigure}
        \caption{  3D RC distribution along the maximum line-of-sight depth gradient, which is perpendicular to the line of nodes (left panels). A positive Z value denotes closer to the observer and a negative value says farther away from the observer (NE is closer and SW farther away from us). Right panels are the edge-on-view of the LMC, where the Y-axis is the distance from the galactic disk plane.  In both the panels, the red dashed line denotes the LMC disk plane and the red solid line traces the median offset from the plane. Bottom panel plots are the same as the top panel plots, but excluding the sub-regions in the central 3$^\circ$ radius.}
        \label{fig002}
\end{figure*}
\subsection{Extinction correction}
The median RC magnitude of each sub-region has to be corrected for the effect of interstellar extinction before we use it for the estimation of the LMC disk parameters. We used the reddening map of the LMC provided by ~\citet{Skowron_2021} to correct for the extinction effect.
The map provides reddening information for different sub-regions in the LMC covering a total area of $\sim$ 180 $deg^2$. For the sub-regions in our study, which are within 6 deg from the LMC centre, we assigned an E(V-I) value of the nearest sub-region in the reddening map. For the sub-regions that are not covered in their study and those which lie beyond 6 deg from the LMC centre, we used the extinction values from \citet{Schlegel_1998}, applying the calibrations given by \citet{2011SchlaflyFinkbeiner}.  
 The extinction in visual band, ($A_v)$ is calculated as, $A_v = 2.352 \times E(V-I)$ as defined by \cite{G_rski_2020} and converted to A$_G$ using a factor of 0.8596, provided 
 by ~\cite{chen2019three}, which was derived using the extinction law from~\citet{cardelli1989relationship}.
 The observed median RC magnitude of each sub-region is corrected for extinction.
\subsection{Distance to each sub-region}
The difference between the extinction corrected RC peak magnitude, ${G_0}$ mag of each sub-region, and the extinction corrected peak magnitude of the LMC centre
is the relative distance modulus ($\mu$) between the LMC centre and each sub-region; $\mu$ is defined as $\mu ={G_0}$ of each sub-region $-$ ${G_0}$ at the LMC centre.  
The relative distance modulus between the LMC centre and each sub-region can then be used to estimate the line of sight distance to each sub-region as: 
$D = D_0 \times 10^{\mu/5} $
, where D$_0$ is the distance to the centre of the LMC which is taken from~\cite{2019Natur.567..200P}, as $49.59 \pm 0.54$ kpc. 
The extinction corrected magnitude (${G_0}$) of the LMC centre (central 10 $\times$ 10 $arcmin^2$ sub-region) is 18.76 mag. In the estimation of the relative distance modulus between each sub-region and the LMC centre, we have assumed a constant intrinsic magnitude for the RC stars across the LMC. However, the intrinsic magnitude of the RC stars across the LMC can vary due to stellar population effects \citep{2016ARA&A..54...95G}. The effect of variation of the intrinsic magnitude of RC stars, due to population effects, on our results are discussed in detail in Section 5.  
\subsection{Plane fit}
The distance (D) and the central coordinates (RA, Dec) of each sub-region are converted to Cartesian coordinates (X, Y, Z) using the transformation equations given in ~\cite{van_der_Marel_2001} and taking 
the photometric centre of the LMC $(\alpha_0, \delta_0) =  (82^\circ.25,-69^\circ.5)$ as the origin. The X-axis is anti-parallel to the right ascension axis, the Y-axis is parallel to the declination axis, and the Z-axis is towards the observer.
We fit a plane to the (X, Y, Z) positions of all the sub-regions by minimising their distances to the plane using an optimisation algorithm (\citealp{KapteynPackage}), where the equation of the plane is defined as Z=AX+BY+C. The best-fit parameters corresponding to the LMC disk plane and the associated errors are calculated. The typical error associated with the z values is $\sim$ 0.03 kpc. 
\begin{figure}
     \begin{subfigure}{0.5\textwidth}
        \includegraphics[width=\textwidth]{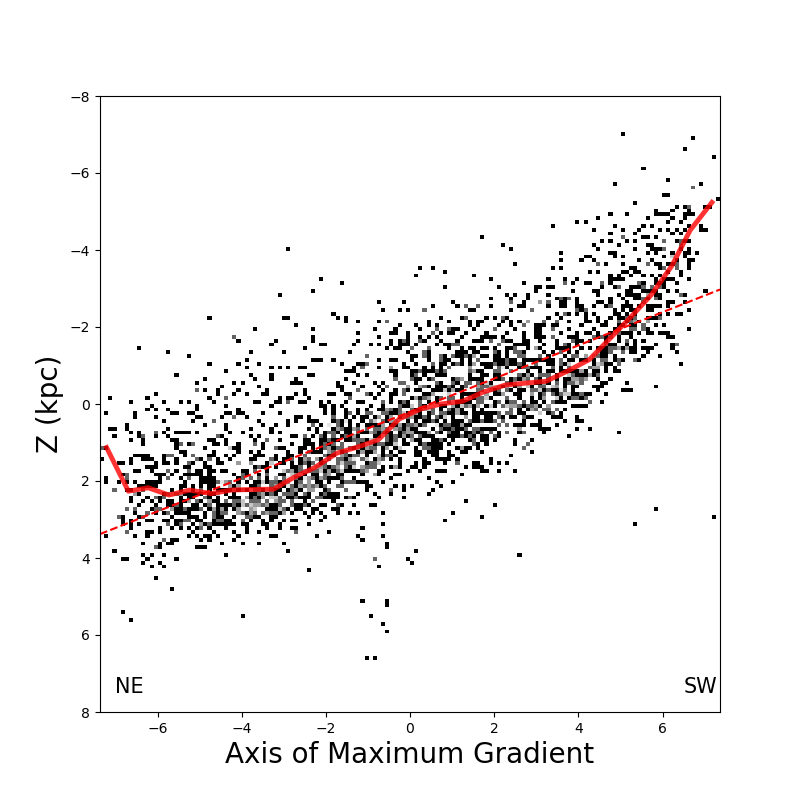}
        \includegraphics[width=\textwidth]{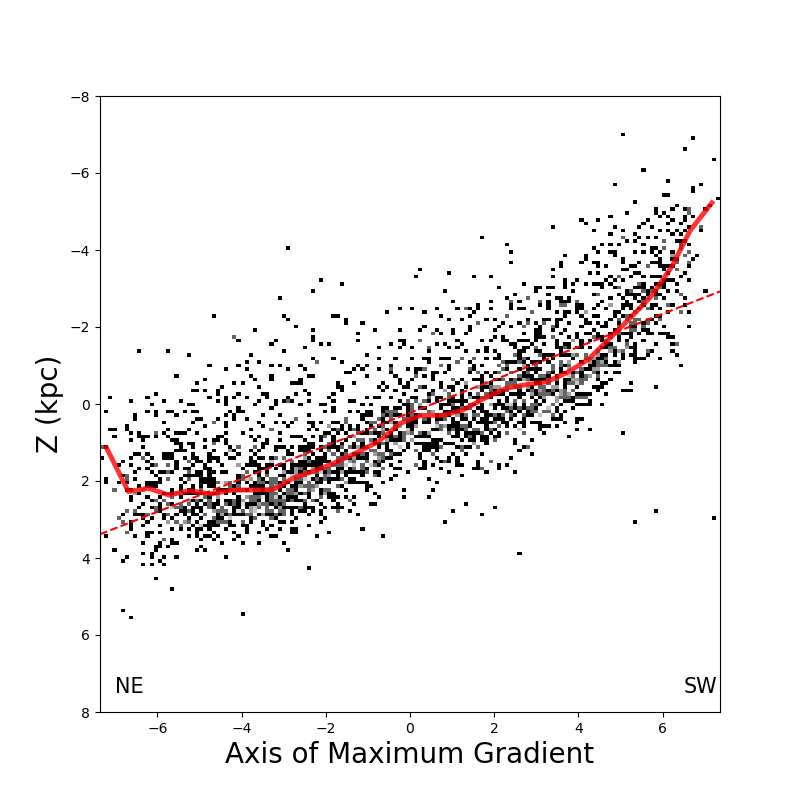}
     \end{subfigure}\\
             \caption{  3D RC distribution along the maximum line-of-sight depth gradient, which is perpendicular to the line of nodes. A positive Z value denotes closer to the observer and a negative value says farther away from the observer (NE is closer and SW farther away from us). Same as the left panels of Figure 3, but all of
the LMC sub-regions are binned with a bin size of 15 $\times$ 15 $arcmin^2$. In the bottom panel, the sub-regions in the central 3$^\circ$ radius are excluded.}
        \label{fig0011}
\end{figure}
\begin{figure*}
     \begin{subfigure}{0.5\textwidth}
         \includegraphics[width=\textwidth]{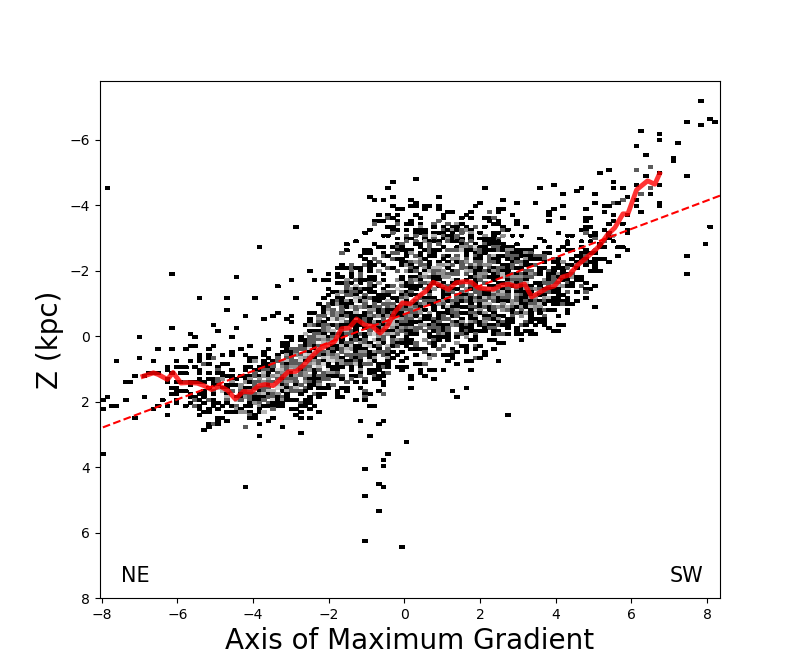}
     \end{subfigure}
     \begin{subfigure}{0.5\textwidth}
        \includegraphics[width=\textwidth]{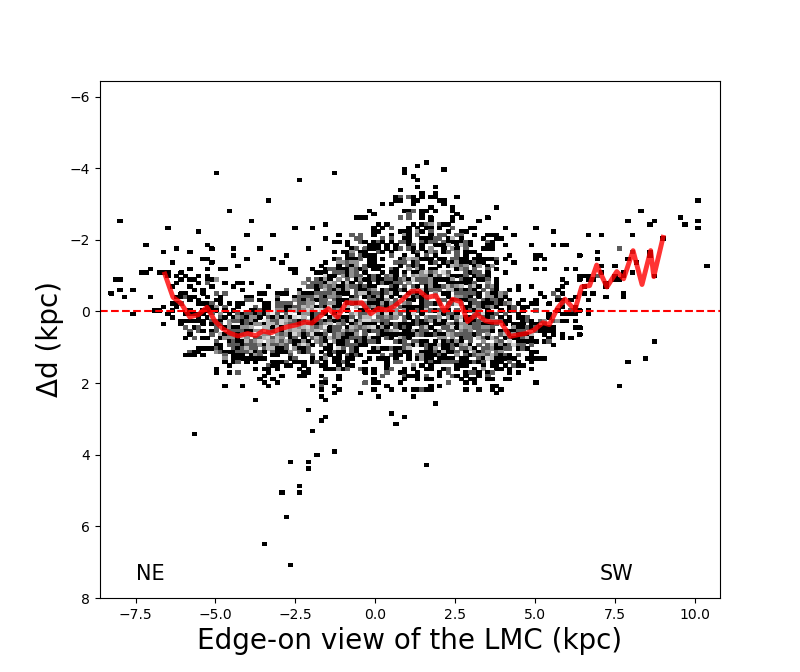}
     \end{subfigure}\\
     \begin{subfigure}{0.5\textwidth}
         \includegraphics[width=\textwidth]{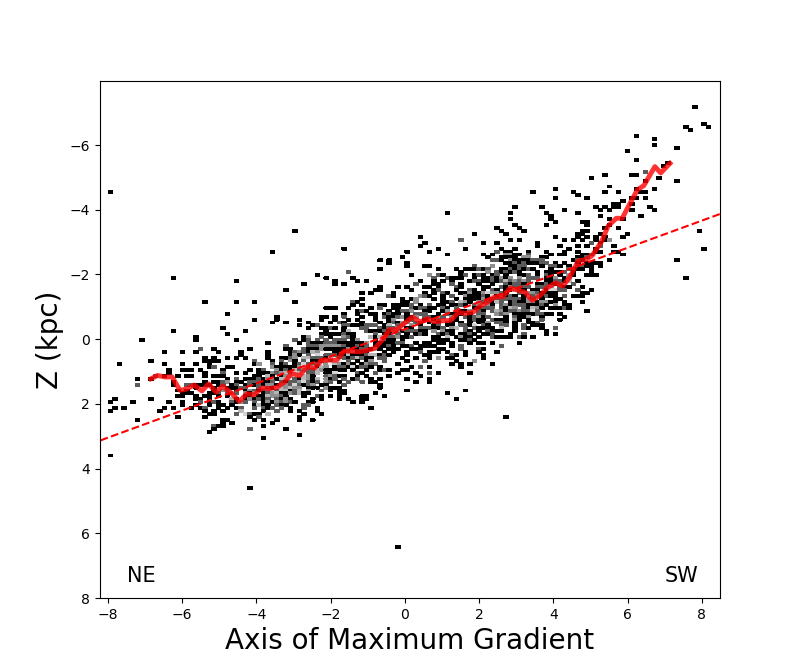}
     \end{subfigure}
     \begin{subfigure}{0.5\textwidth}
        \includegraphics[width=\textwidth]{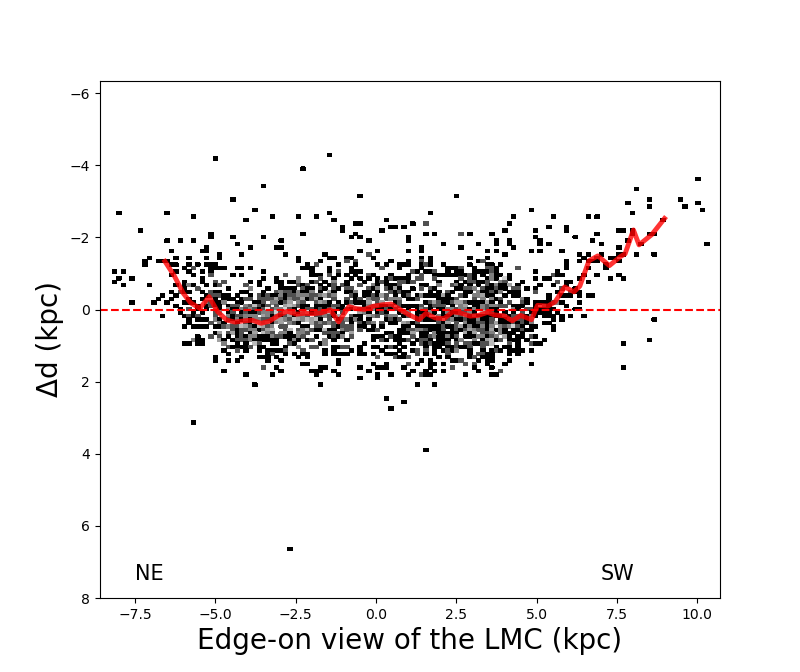}
     \end{subfigure}
         \caption{ 3D RC distribution along the maximum line-of-sight depth gradient, which is perpendicular to the line of nodes. A positive Z value denotes closer to the observer and a negative value says farther away from the observer (NE is closer and SW farther away from us). Same as Figure 3, but including the radial variation of the intrinsic RC magnitude as discussed in \citet{choi2018smashing}.}
        \label{fig004}
\end{figure*}
The best-fit plane parameters are 
$(A, B, C) = ( -0.405 \pm 0.007, 0.144 \pm 0.006, 0.946 \pm 0.017 )$.
We derived the inclination (\textit{i}) and the position angle of the line of nodes $(\theta)$ using the following equations: 
\begin{equation}
 i = cos^{-1}(\frac{1}{\sqrt{A^2+B^2+1}})
,\end{equation}
\begin{equation}
 \theta = tan^{-1}(\frac{-A}{B})+ Sign(B).\frac{\pi}{2}    
.\end{equation}
The uncertainties in \textit{i} and $\theta$ are calculated by propagating the errors of the fit parameters.
For the entire data of $\sim$ 380 $deg^2$ of the LMC disk, the best-fit plane parameters are: $\it{i}$ =  23$^\circ$.26 $\pm$ 0.32  and  $\theta$ = 160$^\circ$.43 $\pm$0.76. These values are comparable with the measurements from previous studies (\citealt{choi2018smashing} and references therein).
\subsection{ Radial dependence of the orientation parameters}
The variation of the LMC disk parameters (\textit{i} \& $\theta$ ) as a function of radius (using the data in different circular and annular regions) is shown in Figure~\ref{fig0033}. Eight annular regions from 2$^\circ$ - 11$^\circ$ radius from the LMC centre are defined. The first seven annular regions have a width of 1$^\circ$ and the last annular region has a width of 2$^\circ$ to have enough sub-regions in the annular region to obtain a reliable estimate of the disk parameters. 
The sub-regions enclosed in the circular region corresponding to the outer radius of each annulus are used to estimate the disk parameters corresponding to each circular region. 
The figure shows that {\it i} \& $\theta$ have a strong radial dependence as observed in earlier studies (\citealp{van_der_Marel_2001,choi2018smashing}).  
In the case of measurements based on the sub-regions enclosed in a circular area, {\it i} decreases from the centre to the outer 4$^\circ$ radius and then remains almost constant. On the other hand, $\theta$ remains nearly constant from the inner to the outer region. In the case of measurements based on fit using sub-regions in different annuli, {\it i} shows a similar trend as that of measurements based on circular region till $\sim$ 4$^\circ$. Then {\it i} increases from 21$^\circ$.85 at 5$^\circ$ radius to 27$^\circ$.3 at 8$^\circ$ and then shows a steep decrease as we move towards the outskirts. A difference in the inclination for the inner and outer disk could be due to the presence of bar in the inner region. This indicates that the bar has a higher inclination than the disk. The value of $\theta$ shows a steep decrease from 163$^\circ$.1 at 5$^\circ$ radius to 150$^\circ$ at 8$^\circ$ and steep increase beyond 8$^\circ$, reaching up to a value of $\sim$ 172$^\circ$ at 11$^\circ$ radius. The sudden variation of {\it i} and $\theta$ at 8--9$^\circ$ annular region 
can be considered as a real variation, since the same variation is seen in the analysis of smaller sub-regions (Section 5.1). The variation indicates that the outer LMC disk is disturbed and might be due to interactions.
Although a similarly strong radial dependence of the disk parameters has been observed in previous studies (\citealt{choi2018smashing,Choi2022LMC-SMCcollision}), the trends are not exactly the same. This could be due to differences in the spatial coverage of the data, especially in the outer LMC. The effect of the size of the sub-region, and hence the number of sub-regions in each annular region, on the radial variation is discussed in Section 5.1.
\section{Possibility of an asymmetric U-shaped stellar warp in the LMC outer disk }
To identify the extra-planar features, the distance to each sub-region is plotted (top left panel of Fig. \ref{fig002}) as a function of the axis of maximum distance gradient (axis perpendicular to the line of nodes).
The LMC disk is inclined with respect to the sky plane in such a way that the NE part is closer to us and the SW part is away from the observer. 
The red solid line in Figure \ref{fig002} shows the median offset of the sub-regions from the best-fit plane, denoted as red dashed line. It shows that some regions of the LMC disk are deviant from the plane. 
In this study, we are mainly interested in the outer regions of the LMC disk. As the inner bar region is found to have a different inclination than the outer disk, we excluded the sub-regions in the central 3$^\circ$ radius region and looked for extra-planar features. The bottom left panel of Figure \ref{fig002} is the same as that of the top-left panel of Figure \ref{fig002}, but excluding the sub-regions in the central 3$^\circ$ radius region. The outer regions clearly show deviations from the best-fit plane. Both the NE and the SW parts of the outer LMC disk are warped away from the observer.\\
The right panels of Figure \ref{fig002} show the edge-on view of the LMC disk, where the Y-axis is the distance of the sub-region from the best-fit plane (a positive value denotes closer to the observer and negative values mean away from the observer). The figures show a new warp in the unexplored NE part of the LMC outer disk. The disk starts deviating away from the LMC best-fit plane at about $\sim$ 5.0 kpc from the centre. The amplitude of the deviation from the best-fit disk plane (as shown by the median offset line in red) is $\sim$ 1.2 kpc. The figure also shows the SW warp (first identified  by~\citealp {choi2018smashing}) which is warped in a direction away from us and extends to $\sim2.4$ kpc away from the best-fit plane, which is lower in amplitude than the findings by \citep{choi2018smashing}. From Figure 3, we can see that both the SW and the NE warps start from $\sim$ 5.0 kpc on either side of the centre and the SW is observed till $\sim$ 9.5 kpc from the centre, whereas the NE warp is seen only till $\sim$ 7.0 kpc from the centre. The NE and the SW warps, together, suggest a probable asymmetric U-shaped warp in the outer LMC disk.\\
The decrease in the amplitude of the SW warp identified in our study compared to that found by \cite{choi2018smashing} could due to multiple factors. One reason could be the difference in the best-fit ($\it{i}$,  $\theta$) values between our study and \cite{choi2018smashing}, which can result in different edge-on view and make the appearance of the stellar warp different. Furthermore, unless the NE and SW warps are perfectly symmetric, a particular edge-on view does not guarantee to show the maximum amplitudes of both warps at the same time. As suggested by \cite{choi2018smashing} and the references therein, the structural parameters of the LMC disk are highly dependent on the spatial coverage of the data used in the analysis. In the present study we use the data from Gaia, which covers the entire LMC disk and, hence, the parameters obtained from this analysis provide a better representation of the LMC disk than in previous studies. The difference in terms of the completeness of the two data sets can also contribute to the observed difference, with SMASH survey being more complete than the Gaia data at RC magnitude range. However, the completeness of sources fainter than 18 mag in G band has significantly improved in the Gaia EDR3 \cite{gaiacollaboration2020gaia}. 
Another reason for the lower amplitude of the SW warp could be our choice of larger area sub-regions in the outskirts, 60 $\times$ 60 $arcmin^2$ or more, compared to 10 $\times$ 10 $arcmin^2$ constant area sub-regions considered by \citep{choi2018smashing}. We chose larger sub-regions in the sparsely populated (due to the applied selection criteria; see Section 2) LMC outskirts to have at least 100-150 stars in the initial RC selection box. Due to the large area sub-regions used, the distances might have averaged out and also the number of data points in the outskirts decreased. The effect of size of the sub-regions and spatial resolution on the newly identified NE warp is discussed in Section 5.1.\\
\section{Discussion}
\subsection{Effect of the size of sub-regions on the stellar warps}
As described in Section 2, the outer regions of the LMC are binned with a larger area, 60 $\times$ 60 and 120 $\times$ 120 $arcmin^2$. Hence, we have a lower number of sub-regions in the region where the stellar warps are identified. To include more sub-regions in our analysis, we binned the entire 11 deg with a bin size of 15 $\times$ 15 $arcmin^2$ and considered only those sub-regions which have at least 25 stars in the redefined RC selection box. A minimum number of 25 stars in the re-defined RC selection box provides detection of probable RC stars with 5$\sigma$ detection. With this criteria, we could get only sub-regions up to 9 deg for further analysis. We performed the same steps and analysis as described in Section 3 to obtain the structural parameters of the LMC disk and to understand the radial dependence of the parameters. We obtained very similar parameters as given in Section 3.5, $i = 23^\circ.18 \pm 0.24,   \theta = 163^\circ.8 \pm 58$, and radial dependence as in Figure 2.\\
To identify the warps, we plotted the distance to each sub-region as a function of the axis of the maximum gradient in Figure 4. The top panel of Figure~\ref{fig0011} shows the entire 9 deg circular region of the LMC from the photometric centre with each point representing a sub-region of 15 $\times$ 15 arcmin$^2$.  The bottom panel is the same as the top panel, excluding the central 3$^\circ$ region of the LMC, containing the bar. The warps in the SW and NE regions are clearly identified as in the top-left and bottom-left panels of Figure 3. As in Figure 3, the NE and the SW warps have an amplitude of $\sim$ 1.2 kpc and 2.5 kpc, respectively,  above the plane and away from the observer. These values are very similar to those obtained in Section 4, from the analysis of sub-regions with different sizes. This result suggests that the size of the sub-region does not significantly affect the estimation of the structural parameters and extra-planar features of the LMC disk. However, we note that there is large scatter in Figure 4 compared to Figure 3 and that could be due to the presence of larger number of sub-regions with a smaller size.
\subsection{Stellar population effects on intrinsic RC magnitude}
In our analysis, we assumed a constant intrinsic RC magnitude across the LMC disk. We note that the intrinsic magnitude of RC stars can vary as a function of age and metallicity \citep{2016ARA&A..54...95G} and, hence, it can vary across the LMC disk. Here, we discuss the effect of population effects on our estimated parameters and the nature of the outer warp. \citet{choi2018smashing} obtained the radial profiles (between 2$^\circ$.7 and 8$^\circ$.5 from the LMC centre) of the intrinsic colour and magnitude of the RC stars in the LMC. These authors calculated  the model predicted RC magnitudes for a given age and metallicity for three radial bins: 2$^\circ$.7 - 4$^\circ$, 4$^\circ$ - 7$^\circ$, and 7$^\circ$ - 8$^\circ$.5, between 2$^\circ$.7 and 8$^\circ$.5. The age and metallicity corresponding to each radial bin was adopted based on the reported metallicity gradient and age-metallicity relation for the LMC disk \citep{Carrera_2008,Carrera_2011,Piatti....145...17P,Pieres/mnras/stw1260}. The adopted metallicity ([Fe/H])  for the three radial bins were -0.5, -0.7, and -0.8. Corresponding ages were 1.6 Gyr, 5.6 Gyr, and 6.3 Gyr, respectively. \\ We computed the mean absolute G band magnitude of  core He-burning stars in these three radial bins using PARSEC isochrones of the corresponding age and metallicity given above, and using Equations (3) and (4) from Girardi \& Salaris (2001). The computed absolute RC magnitude values in the G band for the radial bins, 2$^\circ$.7 - 4$^\circ$, 4$^\circ$ - 7$^\circ$, and 7$^\circ$ - 8$^\circ$.5 are 18.668, 18.696, and 18.693 mag,  respectively. The magnitude in the radial bin 2$^\circ$.7 - 4$^\circ$ is found to be 0.028 mag brighter than the value in 4$^\circ$ - 7$^\circ$ bin. The computed RC magnitudes for the radial bins 4$^\circ$ - 7$^\circ$ and 7$^\circ$ - 8$^\circ$.5 are very similar. The same behaviour was found by \citet{choi2018smashing} for RC magnitudes in {\it i} band. We adopted a gradient of 0.021 mag deg$^{-1}$ between 2$^\circ$.7 - 4$^\circ$ intrinsic RC magnitude and a constant value between 4$^\circ$ - 8$^\circ$.5. For regions inner to  2$^\circ$.7 radius and outer to  8$^\circ$.5 radius, we adopted the absolute RC magnitude at 2$^\circ$.7 and 8$^\circ$.5, respectively. Using the above radial variation for the intrinsic RC magnitude in our analysis, we obtained the inclination and position angle of line of nodes as 23$^\circ.45\pm0.30$ and 159$^\circ.5\pm0.74,$ respectively. The variation of line of sight distance of different sub-regions across the axis of maximum gradient and the edge-on view of the LMC are shown in  Figure \ref{fig004}. The plots are same as in Figure \ref{fig002}, but including the radial variation in the intrinsic RC magnitude due to population effects. The structural parameters and the amplitude of the warps in the NE and SW are very similar to that obtained when the constant intrinsic RC magnitude is used. As observed in Figure 3, the NE warp is better seen when the sub-regions in the inner 3 deg radius are excluded from the analysis.\\
As \citet{choi2018smashing} could not find the complete radial profile of the intrinsic RC magnitude (inside of 2$^\circ$.7 and outside of 8$^\circ$.5 radius), these authors included a gradient for the intrinsic RC magnitude outside 8$^\circ$.5 radius to understand the effect of a lower metallicity and older age RC (considering a [Fe/H] value of $-0.9$ and an age of 10 Gyr at 10$^\circ$.5 radius) population in the outskirts compared to the 7$^\circ$ - 8$^\circ$.5 bin. We computed the absolute G band magnitude  corresponding to a [Fe/H] value of $-0.9$ and an age of 10 Gyr and found it to be 18.798 mag. 
Including a magnitude gradient of 0.053 mag deg$^{-1}$ in the 8$^\circ$.5 - 10$^\circ$.5 radial bin did not change the structural parameters and the overall U-shape of the outer warp remains the same. \\
In nutshell, the above analysis shows that our results on the presence of warp in the NE with a lower amplitude compared to the SW warp and the overall U-shape of the outer LMC warp are not affected significantly by the variation of the intrinsic RC magnitude across the LMC. A recent study by \citet{Mazzi_2021} on the spatially resolved star formation history of the LMC disk using data from VISTA Survey of the Magellanic Clouds (VMC) \cite{cioni2011}, provides a 2D map of the residual of the best-fit plane (left panel of their Figure 13). They consider stars in many other evolutionary stages in their analysis and also their method naturally takes into account the intrinsic variations in the RC absolute magnitude as a function of population age and metallicity.  Though the spatial coverage of the VMC is less compared to our study, their figure clearly shows that the outer NE regions are behind the best-fit plane which is consistent with our results.  
In future, spectroscopic studies of RC stars in the location of the suggested warps will provide more understanding of the effect of stellar population effects on the global shape of the LMC outer warp. 
\begin{figure}
     \begin{subfigure}{0.5\textwidth}
        \includegraphics[width=\textwidth]{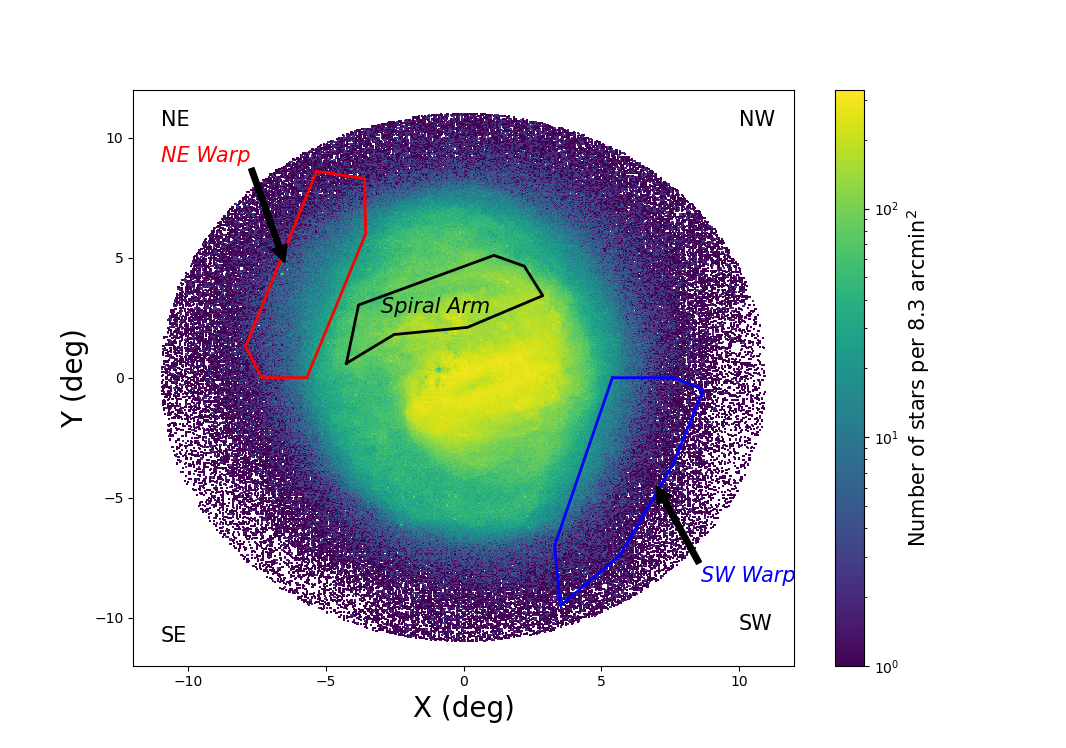}
     \end{subfigure}\\
             \caption{ Cartesian plot with the sources from the \textit Early Data Release 3 (EDR3) for an area of $\sim$ 380 deg$^2$ of the LMC. The colour bar from blue to yellow represents the increase in stellar density (number of stars per bin of size 8.3 $arcmin^2$ , in units of $arcmin^{-2}$ ). The X and Y are defined as in~\citealp{van_der_Marel_2001} with origin as LMC's photometric centre ($82^\circ.25,-69^\circ.45$) ;~\citealp{van_der_Marel_2014}. The east and north are towards left and up, respectively. The approximate locations of the spiral arm (black), north-eastern warp (red), and south-western warp(blue) marked. }
        \label{fig007}
\end{figure}
\subsection{Origin of the LMC outer stellar warp}
Comparing with the modified version of the edge-on view of the LMC stellar density map from the simulations by \cite{2012MNRAS.421.2109B} (Figure. 15 in \citealp{choi2018smashing});~\cite{choi2018smashing} suggested that the observed SW outer warp matches with the simulations. As the SW warp is in the direction towards the SMC, they suggested that the warp might have formed due to the direct collision of the LMC and SMC in the recent past. The simulation plot shows a warp in the NE disk, but in a direction towards the observer and with a lower amplitude. 
Thus an "integral" shaped (S-type) asymmetric warp was expected in the NE outer disk of the LMC. 
The NE warp has a lower amplitude than the SW warp. But in contrary to what is expected, we find a warp in the NE LMC disk which is also bent in the same direction as that of its counterpart in the SW, forming a "U-shaped" asymmetric warp. \\
U-shaped warps are formed mainly due to the super-positions of two or more non-symmetric S-shaped warps \citep{sahajog}. They are caused by multiple fly-bys with different incident angles (\citealp{Kim_2014}). The observed U-shaped warp in the outer LMC disk could have formed due to the multiple interactions between the LMC and the SMC. In the simulations by \cite{2012MNRAS.421.2109B}, the SMC is placed on an eccentric orbit (eccentricity of 0.7) about the LMC and higher orbital eccentricities for the SMC result in fly-by encounters between the MCs. However, this simulation does not show a deviation away from the observer in the NE and instead shows a mild deviation towards the observer. Considering the top-right panel of Figure 3, we see that the median offset line first deviates in the direction towards the observer and then turns away from the observer. Thus, the NE warp launches from below the best-fit plane. This could be the effect of the inclusion of the bar with a different inclination in the estimation of disk structural parameters. In the bottom-right panel of Figure 3, where the sub-regions in the inner 3$^\circ$ are removed, the NE warp is better seen and it launches from the best-fit plane. Thus the presence of bar maybe hiding the real shape of the LMC disk. \citet{2012MNRAS.421.2109B} also indicates that the simulations are not able to fully reproduce all the observed properties of the Magellanic System. As multiple interactions with proper incident angles are required to form U-shaped warps, only a minority of the warps observed are U-shaped warps than the frequently observed S-shaped warps. Thus the observed U-shaped asymmetric warp in the outer LMC disk provides an observational constraint to the theoretical models of the Magellanic system which tries to  understand the LMC-SMC interaction history. We note that other mechanisms like instabilities due to spiral arms can also trigger warp formation. The locations of the warps and the spiral arm of the LMC are shown in Figure \ref{fig007} for reference.
\section{Summary}
Warps are vertical distortions of stellar or gaseous disks of galaxies that are most prominent in less massive galaxies. One of the proposed scenarios for the formation of warps involves tidal interactions among galaxies. In this context, we study the structure of the nearest interacting low-mass disk galaxy, the LMC. 
A recent study by \citet{choi2018smashing} identified a stellar warp in the outer regions of the SW disk of the LMC. However, due to the limited spatial coverage of the data, these authors could not investigate the counterpart of this warp in the NE region, which is essential to understanding the global shape, nature, and origin of the outer LMC warp. In this study, we used the photometric data from \textit{Gaia} EDR3, which covers the entire Magellanic System and we analysed the data of RC stars in the $\sim$ 380 deg$^2$ of the LMC. The \textit{Gaia} EDR3 data cover the entire LMC disk, including the very outskirts of the galaxy. The entire data were binned into a total of 3626 sub-regions having enough stellar density to visually identify the RC stars. The extinction corrected peak RC magnitude were used to estimate the orientation parameters of the LMC disk plane. For the $11^\circ$ radial region the LMC disk, the inclination (\textit{i}), and position angle of the line of nodes ($\theta$) are found to be, $\it{i}$ =  23$^\circ$.26 $\pm$ 0.32,  and  $\theta$ = 160$^\circ$.43 $\pm$0.76 and our results are consistent with the previously estimated values. 
For the first time, we detected a warp in the NE outer disk of the LMC with an amplitude $\sim1.2$ kpc in a direction away from us. We also detected the warp identified by~\cite
{choi2018smashing} in the SW part. Both warps are bent in the same direction, suggesting an asymmetric U-shaped warp. Our result provides an observational constraint to the theoretical models of the Magellanic system, which are aimed at attaining an understanding of the LMC-SMC interaction history.\\\\
{\bf Acknowledgements:}
    Saroon S acknowledges support from the Indian Institute of Astrophysics through the Visiting Students Programme. Smitha Subramanian acknowledges support from the Science and Engineering Research Board of India through a Ramanujan Fellowship. This work has made use of data from the European Space Agency (ESA) space mission Gaia (\url{https://www.cosmos.esa.int/gaia}). Gaia data are being processed by the Gaia Data Processing and Analysis Consortium (DPAC). Funding for the DPAC is provided by national institutions, in particular the institutions participating in the Gaia MultiLateral Agreement (MLA). This research made use of numpy \citep{numpy}, scipy \citep{scipy2020}, matplotlib \citep{matplotlib}, lmfit (\citealp{2016ascl.soft06014N}) ,kmpfit (\citealp{KapteynPackage}) and  astropy\footnote{http://www.astropy.org}, a community-developed core Python package for Astronomy \citep{astropy:2013, astropy:2018}. Finally, it is a pleasure to thank the referee for a constructive report. \\
\bibliography{biblio.bib}
\end{document}